\documentclass[11.7ptt]{article}

\usepackage{amsmath}
\usepackage{amssymb}
\def\b{\begin{equation}}
\def\e{\end{equation}}

\def\b{\begin{equation}}
\def\e{\end{equation}}
\def\bd{\begin{displaystyle}}
\def\ed{\end{displaystyle}}
\def\ba{\begin{array}}
\def\ea{\end{array}}

\def\bee{\begin{enumerate}}
\def\eee{\end{enumerate}}

\def\bes{\begin{eqnarray*}}
\def\ees{\end{eqnarray*}}
\def\be{\begin{eqnarray}}
\def\ee{\end{eqnarray}}

\begin{document}
\title{Radiation effects on particle's trajectory in the linear level}

\author{M. Fathi$^1$,
 M. Tanhayi-Ahari$^2$, M.R. Tanhayi$^1$
 \thanks{e-mail: m$_{-}$tanhayi@iauctb.ac.ir}, F. Tavakoli$^1$\thanks{e-mail: f$_{-}$tavakoli@iauctb.ac.ir}}
\maketitle   \centerline{\it $^{1}$Department of Physics, Islamic
Azad University, Central Tehran Branch, Tehran, Iran}
\centerline{\it$^2$ Department of Physics, Sharif University of
Technology (SUT), Tehran, Iran}

\begin{abstract}

In this work, we first obtain the linear form of the scalar
self-force and then, the effect of self-force on the particle's
trajectory is considered. In de Sitter space-time, within the
classical approach, we consider this effect. Finally, some limits
for the problem are presented.

\end{abstract}

\section{Introduction}

In many cases of physical interest, for example in cosmology,
radiation is an important subject, since almost all our measurable
knowledge of universe is based on radiation. It is well known that
radiation of an accelerated charged particle affects the motion of
the particle which is known as the Abraham-Lorentz force
\cite{abraham}. The Abraham-Lorentz force is the recoil force that
acts back on the radiating particle. It is also called the
radiation reaction force. This force carries momentum and is
proportional to the square of the charge and also the rate of
change in acceleration or the "jerk". Dirac \cite{dirac} employed
space-like geodesics to generalize the Abraham-Lorentz force to
the relativistic velocities. Hence, this self-force which is due
to the particle's own electromagnetic field on itself is called
the Abraham-Lorentz-Dirac force. Moreover, in physics we deal with
particle as a point-like object. Then an immediate question
arises: What happens to the particle's energy on exactly the
particle's position? clearly it diverges at the particle's world
line. Such difficulty is discussed and resolved by the mean of
self-energy \cite{gralla}, and one can find a lot of papers on
these subjects (for further review see \cite{poisson} and ref.s
there in).

In the present study, we limit our analysis to the self-force
effect upon the trajectory of an extended charged particle in the
linear level in flat background and also in non-relativistic limit
in de Sitter space-time. We choose the de Sitter space-time,
because according to the cosmological data, this model could well
describe our universe with a small non-vanishing cosmological
constant \cite{riess}. In this space-time, we obtain the geodesic
equations for an extended charged particle. Then by solving these
equations, the relation for the particle's velocity with respect
to time is obtained. This relation will be used to derive the
radiation reaction for a non-relativistic charged particle,
falling freely in a gravitational field defined by de Sitter
geometry.

The organization of this paper is as follows: in section two, we
review the self-force, and then we obtain the linear form of the
scalar self-force. After that, in flat background, the equation of
motion is analyzed. In section 3, after reviewing de Sitter
space-time, the geodesics for a freely falling particle in this
space-time is studied and as it is shown one can find a
gravitational force acting on this particle in local
consideration. In this space-time, we find the corrections in the
geodesics due to the self-force in the non-relativistic limit.

 In this paper for the sake of simplicity, we set $c=1$.

\section{Self-force}

The self-force on a charge arises from the interaction of the
charge with its own retarded field. The resultant acceleration is
proportional to $q^2/m$ and the covariant form of the self-force
is given by \cite{rohrlich}
\begin{equation}
F_{self}^\mu=F^\mu_{Sch}+F^\mu_{rad},\label{j1}
\end{equation}
where $F^\mu_{Sch}$ and $F^\mu_{rad}$ are the Schott and radiation
reaction terms respectively which are defined by:
\begin{equation} F^\mu_{Sch}=\frac{2}{3}{q^2}\ddot u^\mu,\,\,\,\,
F^\mu_{rad}= \frac{2}{3}{q^2}u^\mu \dot u^\lambda \dot
u_\lambda,
\end{equation}
where dot stands for the time derivative. Thus the Lorentz-Dirac
equation reads:
\begin{equation} \label{ld}
ma^\mu= F^\mu_{ext}+F^\mu_{self}.\end{equation} This equation is a
third-order differential equation and hence in the consideration
of the equation of motion, some difficulties arise, for example,
the appearance of the runaway solutions. It means, with a constant
applied force the acceleration grows exponentially with time
\cite{gralla, poisson}. Some authors explore the quantum
mechanical response to these problems \cite{moniz}, but
Landau-Lifshitz's remedy seems to work well, in which they
considered ${q^2}/{m}$ as a very small quantity. This condition
means that the radiation reaction must be very small in comparison
with the external force that acts on the charge
\cite{poisson,spirin, landau}, so we have:
\begin{equation}\label{self1}
ma^\mu=F_{ext}^\mu+\frac{2}{3}\frac{q^2}{m}\Big(\delta_\lambda^\mu+u^\mu
u_\lambda\Big)\partial_\gamma
F^\lambda_{ext}u^\gamma,\end{equation} and its non-relativistic
limit is as follows \cite{poisson, landau}:
\begin{equation} \label{nr}
ma=f_{ext}+\frac{2}{3}\frac{q^2}{m}\dot f_{ext}.
\end{equation}
The above formulas are in the flat geometry and the analysis of
the electromagnetic and gravitational self-forces in curved
space-time were first considered in \cite{dewitt} and \cite{mino},
respectively. But what we are going to consider in this paper is
the self-force acting on a particle with scalar charge $q$ in
curved space-time with the following equation of motion
\cite{quinn}:

\begin{equation} \label{qui} m
\frac{D}{d\tau} u^\mu
=f^\mu_{ext}+\frac{1}{3}\frac{q^2}{m}(\delta^{\mu}_{\nu}+u^\mu
u_\nu)\dot{f}^\nu_{ext}+\frac{1}{6}q^2(R^{\mu}_{\nu} u^\nu+u^\mu
R_{\nu\gamma} u^\nu u^\gamma)+f^\mu_{self},
\end{equation} where
$\frac{D}{d\tau}u_\mu =u^\nu\nabla_\nu u_\mu$ defines the
acceleration, $R_{\mu\nu}$ is the Ricci tensor and the self force
is given by
\begin{equation} \label{self} f^\mu_{self}=q^2\int_{-\infty}^{\tau}\Big(\partial^\nu
G_R(x,x')+u^\mu u_\nu\partial^\nu
G_R(x,x')\Big)\,d\tau',
\end{equation} $G_R(x,x')$ is the retarded
scalar Green's function in that space-time. This is the simplest
generalization of the Eq. (\ref{ld}) and it recovers Eq.
(\ref{self1}) in the flat space-time \cite{poisson1}. By the
scalar charge, it means that the charge q is constant along the
world line of the particle, note that in contrast to the
electromagnetic case, the Klein-Gordon equation does not require
conservation of charge \cite{quinn}.

\subsection{Self-force at the linear level}

At this stage let us consider Eq. (\ref{qui}) in the weak field
approximation. Suppose that we have a small deviation from the
background as:
$$g_{\mu\nu}=\overline{g}_{\mu\nu}+h_{\mu\nu},$$
where $\overline{g}_{\alpha\beta}$ is the background metric and
$h_{\alpha\beta}$ is very small compared to
$\overline{g}_{\alpha\beta}$ at every point. It is shown that the
linear part of Christoffel connections, Riemann tensor and Ricci
tensor become as follows \cite{gullu}:

\begin{equation}
(\Gamma^\alpha_{\mu\rho})_L=\frac{1}{2}\overline{g}^{\alpha\lambda}(\overline{\nabla}_\mu
h_{\rho\lambda}+\overline{\nabla}_\rho
h_{\mu\lambda}-\overline{\nabla}_\lambda
h_{\mu\rho}),\end{equation}
\begin{equation}
(R^\mu\,_{\nu\rho\sigma})_L=\frac{1}{2}\Big(\bar{\nabla}_\rho\bar{\nabla}_\sigma
h^\mu_\nu+\bar{\nabla}_\rho\bar{\nabla}_\nu
h^\mu_\sigma-\bar{\nabla}_\rho\bar{\nabla}^\mu
h_{\sigma\nu}-\bar{\nabla}_\sigma\bar{\nabla}_\rho
h^\mu_\nu-\bar{\nabla}_\sigma\bar{\nabla}_\nu
h^\mu_\rho+\bar{\nabla}_\sigma\bar{\nabla}^\mu
h_{\rho\nu}\Big),\end{equation} \begin{equation}
(R_{\nu\sigma})_L=\frac{1}{2}\Big(\bar{\nabla}_\mu\bar{\nabla}_\sigma
h^\mu_\nu+\bar{\nabla}_\mu\bar{\nabla}_\nu
h^\mu_\sigma-\bar{\Box}h_{\sigma\nu}-\bar{\nabla}_\sigma\bar{\nabla}_\nu
h\Big),\end{equation} \b R_L=\bar{g}^{\alpha\beta}
(R_{\alpha\beta})_L-\bar{R}^{\alpha\beta}h_{\alpha\beta},\e \b
u^\mu=\bar{u}^\mu+(\Gamma_{0\lambda}^\mu)_L x^\lambda.\e

With these relations in hand, let us consider the linear form Eq.
(\ref{qui}) in generic background without any external force.
First order expansion of the left hand side of (\ref{qui})
becomes:
\begin{equation}
\begin{aligned} m(u^\nu\nabla_\nu u^\mu)_L=m\bar{u}^\nu\bar{\nabla}_\nu\Big((\Gamma_{0\lambda}^\mu)_L\,x^\lambda\Big)+
m \bar{u}^\nu(\Gamma_{\lambda\nu}^\mu)_L\bar{u}^
\lambda+m(\bar{\nabla}_\nu\bar{u}^\mu)(\Gamma_{0\lambda}^\nu)_L\,x^\lambda.
\end{aligned}
\end{equation}
At the right hand side we have $R^{\mu}_{\nu}u^\nu$ and $u^\mu
R_{\nu\gamma}u^\nu u^\gamma$, where up to first order after doing
some calculation, can be written as follow: \begin{equation}
\begin{aligned}
(R^\mu_{\nu}u^\nu)_L&=\bar{g}^{\mu\alpha}({R}_{\alpha\nu})_L\bar{u}^\nu-h^{\mu\alpha}\bar{R}_{\alpha\nu}\bar{u}^\nu+
\bar{R}^\mu_\nu(\Gamma_{0\lambda}^\nu)_L\,x^\lambda,\\
(u^\mu {R}_{\nu\gamma} u^\nu u^\gamma)_L&=\bar{u}^\mu\bar{u}^\nu
\bar{u}^\gamma(R_{\nu\gamma})_L+
\bar{R}_{\nu\gamma}\bar{u}^\nu\bar{u}^\gamma(\Gamma_{0\lambda}^\mu)_L\,x^\lambda
+\bar{R}_{\nu\gamma}\bar{u}^\gamma\bar{u}^\mu(\Gamma_{0\lambda}^\nu)_L\,x^\lambda\\
&+\bar{u}^\mu\bar{R}_{\nu\gamma}\bar{u}^\nu(\Gamma_{0\lambda}^\gamma)_L\,x^\lambda,
\end{aligned}
\end{equation}
Therefore at the linear level without any external forces,
equation (\ref{qui}) turns to:
\begin{equation}
\begin{aligned}\label{lineareq}
m\bigg
\{\bar{u}^\nu\bar{\nabla}_\nu\Big(&(\Gamma_{0\lambda}^\mu)_L\,x^\lambda\Big)+
\bar{u}^\nu(\Gamma_{\lambda\nu}^\mu)_L\bar{u}^
\lambda+(\bar{\nabla}_\nu\bar{u}^\mu)(\Gamma_{0\lambda}^\nu)_L\,x^\lambda\bigg\}=
\frac{q^2}{6}\bigg\{\bar{g}^{\mu\alpha}({R}_{\alpha\nu})_L\bar{u}^\nu
-h^{\mu\alpha}\bar{R}_{\alpha\nu}\bar{u}^\nu\\
&+\bar{R}^\mu_\nu(\Gamma_{0\lambda}^\nu)_L\,x^\lambda+\bar{u}^\mu\bar{u}^\nu
\bar{u}^\gamma(R_{\nu\gamma})_L+
\bar{R}_{\nu\gamma}\bar{u}^\nu\bar{u}^\gamma(\Gamma_{0\lambda}^\mu)_L\,x^\lambda
+\bar{R}_{\nu\gamma}\bar{u}^\gamma\bar{u}^\mu(\Gamma_{0\lambda}^\nu)_L\,x^\lambda\\
&+\bar{u}^\mu\bar{R}_{\nu\gamma}\bar{u}^\nu(\Gamma_{0\lambda}^\gamma)_L\,x^\lambda\bigg\}+\Big(f^\mu_{self}\Big)_L.
\end{aligned}
\end{equation}
Where the linear form of the self-force is as follows:
\begin{equation}
\begin{aligned}
\Big(f^\mu_{self}\Big)_L=&q^2 \Big(2\bar{u}^\mu\bar{u}^\alpha
h_{\nu\alpha}+\bar{g}_{\nu\alpha}\bar{u}^\mu(\Gamma_{0\lambda}^\alpha)_L\,x^\lambda
+\bar{g}_{\nu\alpha}\bar{u}^\alpha(\Gamma_{0\lambda}^\mu)_L\,x^\lambda-h^{\mu}_\nu\Big)\int_{-\infty}^\tau\partial^\nu
G_R(x,x')d\tau'.
\end{aligned}
\end{equation}
The equation (\ref{lineareq}) can be considered as the linear form
of (\ref{qui}) in generic background. What one shall do is
computing the linear Christoffel connections, Riemann and Ricci
tensors and also the proper Green's functions.

\subsection {Flat background case:}

Suppose in flat background we have a small perturbation, $h_{\mu\nu}$, where  \b \label{per} h_{\mu\nu}=\left\{%
\begin{array}{ll}
    h_{00}=-\phi(r),  \\
    h_{ij}=-\phi(r), & i=j \\
\end{array}%
\right. \e  where $\phi(r)$ is a time-independent scalar function
of $r$. And also 4-velocity in flat background has been used as
$\bar{u}^\mu=\gamma(1,\overrightarrow{v})$, where $\gamma$ is the
usual relativity factor, but in this case we consider it as
$\bar{u}^\mu=(1,0,0,0)$, namely for the comoving observer while
relates to the other inertial observers by the Lorentz
transformation. Note that hereafter we omit the bar over the
velocities since they are all written in the background. It is
easy to calculate that
\begin{equation}
\begin{aligned}
(\Gamma_{0\lambda}^\mu)_L\,x^\lambda&=\frac{1}{2}\eta^{\mu\alpha}(r\partial_rh_{0\alpha}-t\partial_\alpha h_{00}),\\
(\Gamma_{0\lambda}^\nu)_L\,x^\lambda\partial_\nu{u}^\mu&=-\frac{1}{2}\partial_rh_{00}(r\partial_0+t\partial_r){u}^\mu,\\
(R^\mu_\nu u^\nu)_L &= \eta^{\mu\alpha}(R_{\alpha\nu})_L{u}^\nu
\end{aligned}
\end{equation}
After doing some calculation we obtain
\begin{equation}
\begin{aligned}\label{eqw}
\partial_0(r\partial_r h_0^\mu-t\partial^\mu h_{00})-\partial^\mu
h_{00}&=-\frac{q^2}{6m}\Big(\Box h_0^\mu+u^\mu\Box
h_{00}\Big)+\frac{q^2}{m}\Big(4u^\mu h_{0\nu}+u^\mu(r\partial_r
h_{0\nu}-t\partial_\nu h_{00})\\
&+u_\nu(r\partial_r h_0^\mu-t\partial^\mu
h_{00})-2h_\nu^\mu\Big)\int_{-\infty}^\tau\partial^\nu
G_R(x,x')d\tau',
\end{aligned}
\end{equation}
we have maintained $u^\mu$ in order to keep the covariance. The
equation (\ref{eqw}) can be simplified as:
\begin{itemize}
    \item for $\mu=0$, we obtain:
\b\label{eqw1}
\dot{r}\partial_r\phi=-\frac{q^2}{m}\Big((6\phi+2r\partial_r\phi)\partial_0
{\cal A}+t\partial_r\phi\partial_r{\cal A}\Big),\e
    \item for $\mu=1$, we have:
    \b \label{eqw2}
    2\partial_r\phi=\frac{q^2}{m}\Big(t\partial_r\phi\partial_0{\cal
    A}+2\phi\partial_r{\cal A}\Big), \e
    Where ${\cal A}\equiv\int_{-\infty}^\tau
G_R(x,x')d\tau'$.
\end{itemize}
From (\ref{eqw1}) and (\ref{eqw2}) we obtain:

\b 2\dot{r}\partial_r{\cal
A}=\frac{q^2}{m}\Big[(6t-\frac{12m}{q^2}-4r\partial_r{\cal
A})\partial_0{\cal A}-2(\partial_r{\cal A})^2\Big]. \e

What is interesting here, can be stated as follows: \\when we
consider the perturbation in the flat background, the trajectory
of the particle is independent of the perturbation factor, at
least for our case when $\phi$ is supposed to be a function of $r$
only.

Now let us recall the Green's function in the flat space-time.
\subsection{Green's functions:}
Green's functions are actually an integral kernel that can be used
to solve an inhomogeneous differential equation with boundary
conditions in which for a complete set of modes they are defined
by
$$G(x,x')=\sum_n\phi_n(x)^\ast\phi_n(x').$$
In quantum mechanics, the vacuum expectation values of various
products of free field operators can be identified with various
Green's functions of the wave equation. For example, the
Pauli-Jordan or Schwinger function is defined by the vacuum
expectation value of the commutator of fields as follows
\cite{birrell} \b iG(x,x')=\langle
0|(\phi(x)\phi(x')-\phi(x')\phi(x))|0\rangle.\e

One can define retarded and advance Green's functions, which are
relevant to our work, as
\begin{equation}
\begin{aligned}
G_R(x,x')&=-\theta(t-t')G(x,x'),\\
G_A(x,x')&=\theta(t'-t)G(x,x'),
\end{aligned}
\end{equation}
where \b \theta(t)=\left\{%
\begin{array}{ll}
    1, & t>0; \\
    0, & t<0. \\
\end{array}%
\right. \e Note that as mentioned all these forms satisfy in field
equation
$$ (\Box-m^2)G_{A,R}(x,x')=\delta^4(x,x').$$
In flat space the proper retarded Green's function is given by \b
G_R(\overrightarrow{x},t;\overrightarrow{x}',t')=\frac{\delta(t-t'+|\overrightarrow{x}-\overrightarrow{x}'|)}{|\overrightarrow{x}-\overrightarrow{x}'|},
\e

\section{de Sitter Space-time}

Recent observational data are strongly in favor of a positive
acceleration of our expanding universe \cite{riess}. In the first
approximation, the background space-time might be considered as de
Sitter space-time. The de Sitter space-time is a solution of the
vacuum Einstein equation with a positive cosmological constant
$\Lambda$,
\begin{equation}
\begin{array}{l}
 R_{\mu\nu}- \frac{1}{2} R g_{\mu\nu} + \Lambda g_{\mu\nu} = 0.
 \end{array}
 \label{Enstein equation}
 \end{equation}
The de Sitter space-time is the unique maximally symmetric curved
space-time with ten Killing vectors (the same as Minkowski
space-time) and locally characterized by the condition
\cite{hawking}:
\begin{equation}
\begin{array}{l}
\bar{R}_{\mu\nu\lambda\rho} = \frac{2\Lambda}{(D-1)(D-2)}
(\bar{g}_{\mu\lambda}\bar{g}_{\nu\rho}-
\bar{g}_{\mu\rho}\bar{g}_{\nu\lambda}).
 \end{array}
 \label{Riemann tensor}
 \end{equation}
 $\bar{R}_{\mu\nu\lambda\rho}$ is the Riemann curvature tensor. Using
the relations $\bar{R}_{\mu\nu} =
\bar{R}^{\lambda}_{\mu\lambda\nu}$ , $\bar{R} = \bar{g}_{\mu\nu}
\bar{R}^{\mu\nu}$ we obtain:
\begin{equation}
\begin{array}{l}
\bar{R}_{\mu\nu}=\frac{2\Lambda}{D-2}\bar{g}_{\mu\nu},\\\\

 \bar{R} =
\frac{2D}{D-2}\Lambda,
\end{array}
 \label{Ricci scalar}
 \end{equation}
where $R$ is the Ricci scalar and $D$ is the dimension of the
space-time. The metric in de Sitter space-time is defined by
\begin{equation}
\begin{array}{l}
ds^2 = \bar{g}_{\mu\nu} dX^\mu dX^\nu, \quad\mu , \nu =1,2,\cdots
D,
\end{array}
 \label{de sitter metric general}
 \end{equation}
in which $X^\mu$ belongs to the intrinsic coordinates of this four
dimensional hyperbolic space-time. Note that for the sake of
simplicity, only in this section, the Greek indexes run from $1$.
It means that de Sitter space-time can be represented by a
hyperboloid, embedded in a $D+1$-dimensional flat space (Ambient
space notation), with one constraint:
\begin{equation}
\begin{array}{l}
X_H = \{x \in R^{D+1} ; x^2 = \eta_{ab}x^a x^b = H^{-2}\},\quad
 a ,\ b =1,2,\cdots ,D+1,
 \end{array}
 \label{X_H}
 \end{equation}
 for which the metric is given by:
\begin{equation}
\begin{array}{l}
$$ds^2 = \eta_{ab} dx^a dx^b, \quad
 \eta_{ab} = diag\{-1,1,1,\cdots ,(D+1)\},
\end{array}
\label{de Sitter metric - 5D}
\end{equation}
where $H^{- 1}$ is the minimum radius of the hyperboloid and $H$
is the Hubble parameter, where we have \b
\Lambda=\frac{(D-2)(D-1)}{2}H^2.\e In this work we take $D=4$.

Generally, one can define four classes of coordinate systems in de
Sitter space-time, Global, Conformal, Flat and Static (for more
review see \cite{spradlin}). Here, we recall two of them, Flat and
Static coordinates:

\begin{itemize}

\item[i)] Flat coordinates: This metric is defined
by:\begin{equation}\label{de Sitter flat metric} ds^2=\eta_{ab}
dx^a dx^b=\bar{g}_{\mu\nu}dX^\mu dX^\nu=\-dt^2 +
e^{-2Ht}dX_i^2,\,\,\,\,\,\,i=2,3,4\end{equation} where
\begin{equation}
\begin{aligned}
x_1=&H^{-1}\sinh Ht-\frac{1}{2}He^{-Ht}X_iX_i, \\
x_i=&e^{-Ht}X_i,\\
x_5=&H^{-1}\cosh Ht-\frac{1}{2}He^{-Ht}X_iX^i.
\end{aligned}
\end{equation}
This metric due to its simplicity and similarity to the Minkowski
space-time, is more applicable in literature, note that the
spacial part is flat.

\item[ii)] Static coordinates: The metric is defined by
\begin{equation} \label{static}
ds^2=-(1-H^2r^2)dt^2+(1-H^2r^2)^{-1}dr^2+r^2d\Omega^2,\end{equation}
where
\begin{equation}
\begin{aligned}
x_1=&H\sqrt{1-H^2r^2}\sinh Ht, \\
x_2=&r\sin\theta\cos\varphi,\\
x_3=&r\sin\theta\sin\varphi,\\
x_4=&r\cos\theta,\\
x_5=&H\sqrt{1-H^2r^2}\cosh Ht.
\end{aligned}
\end{equation}

Only in this coordinate system $\frac{\partial}{\partial t}$ is a
time-like Killing vector, and due to the this property it is
frequently used by authors. We leave more details to the relevant
references.
\end{itemize}

It is worth to mention that although de Sitter space-time has a
non-vanishing Ricci scalar, however in local consideration and in
the mentioned coordinate, Eq. (\ref{nr}) is still applicable.
Therefore, the modified equation of motion can be considered by
replacing $f_{total}\equiv ma$ in the total force and
$f_{self}=\frac{2}{3}{q^2}\ddot v$. It is easy to show that the
following relation holds between $f_{self}$ and $f_{ext}$ (see
appendix):
\begin{equation}
f_{self}=\sum^{\infty}_{n=1}\alpha^n(\frac{d^n}{dt^n})f_{ext},
\label{E15}
\end{equation}
where $\alpha \equiv \frac{2}{3}\frac{ q^2}{mc^3}.$ Here, we
consider only the first derivative, because $\frac{q^2}{m}$ is
considered to be very small. So one obtains:
\begin{equation}
f_{total}=\sum^{\infty}_{n=0}\alpha^n(\frac{d^n}{dt^n})f_{ext}.
\label{E16}
\end{equation}

\subsection{Geodesics:}

The aim of this section is to obtain the geodesic equations in de
Sitter space-time. These equations can be deduced from
\begin{equation}
\begin{array}{l}
\frac{d^2X^{\mu}}{d\tau^2} + \Gamma^{\mu}_{\nu\rho}
\frac{dX^\nu}{d\tau} \frac{dX^\rho}{d \tau} = 0,
\end{array}
 \label{Geodesic equations formula}
 \end{equation}
where $\tau$ is the affine parameter that here is considered as
the proper time. Using the metric (\ref{de Sitter flat metric}) in
Eq. (\ref{Geodesic equations formula}) gives us the geodesic
equations:
$${\frac {d^{2}}{d{\tau}^{2}}}t \left( \tau \right)
=-H{ {\rm e}^{2\,Ht}}\sum_{i=1}^{3}{ \left( {\frac
{d}{d\tau}}X_{{i}} \left( \tau \right)
 \right) ^{2}},$$
\begin{equation}
\begin{array}{l}
{\frac {d^{2}}{d{\tau}^{2}}}X_{{i}} \left( \tau \right) =-2\,H
\left( {\frac {d}{d\tau}}t \left( \tau \right)  \right) {\frac
{d}{d\tau}}X_{ {i}} \left( \tau \right).
\end{array}
 \label{Geodesic equations}
 \end{equation}
Let us consider only the $i=1$ case. From the geodesic equations,
we can obtain two differential equations:
\begin{equation}
\begin{array}{l}
\frac{d}{d\tau}X(\tau)=e^{-2Ht(\tau)}
 \end{array}
 \label{e1}
 \end{equation}
\begin{equation}
\begin{array}{l}
\frac{d^2}{d\tau^2}t(\tau)=-{3}He^{-2Ht(\tau)}.
 \end{array}
 \label{e2}
 \end{equation}
After doing some algebra we obtain (appendix)
\begin{equation}
\dot v=-2Hv\Big(1-\frac{3}{2}{v^2}e^{2Ht}\Big), \label{dotv}
\end{equation}
where $v=\dot{X}$. Due to the geodesic equations, any rest mass in
a local frame, is accelerating, and then the four-acceleration
$a^\mu$ is non-zero. This was shown in equation (\ref{dotv}),
which confirms that the massive object is being pushed off its
geodesics, and therefore is accelerating. Note that if we suppose
${v}\ll 1$, then we will obtain $v=v_0e^{-2Ht}$; this formula
notifies the motion of an object in a fluid. From Eq.
(\ref{dotv}), we obtain:
\begin{equation}
v=\frac{e^{-Ht}}{\sqrt{3+e^{2Ht}\Big(\frac{1}{v_0^2}-3\Big)}}.
\label{v}
\end{equation}
Thus the following condition on the initial velocity must be
satisfied:
$$v_0<\frac{1}{\sqrt{3(1-e^{-2Ht})}}.$$

In the flat limit case ($H\longrightarrow 0$), we have no change
in particle's velocity. It means that the curvature change the
velocity as we expected. The case $v_0=\frac{1}{\sqrt{3}} $, is
somehow interesting since in the weak field limit, a particle
moving radially in Schwarzschild background, neither accelerates
nor decelerates\footnote{$v=\frac{1}{\sqrt{3}}$ is also the speed
of sound in an ultra-relativistic medium, note that we have taken
$c=1$.} \cite{blinnikov}. With this initial velocity, we obtain \b
v=\frac{e^{-Ht}}{\sqrt{3}},\e on the other hand for a particle
that initiated its motion from the early universe, we can write
$H_0T\approx1$ \cite{data} - where $T$ stands for the present age
of our universe - so the initial velocity in the early universe
should have been less than $0.7$.
\\Finally one can obtain the following statement for the
trajectory of the particle in de Sitter universe:
\begin{equation}
X(t)=\frac{1}{3H}\Big(\frac{1}{v_0}-e^{-Ht}\sqrt{3+(-3+\frac{1}{v_0^2})e^{2Ht}}\,\,\Big),
\label{tr}
\end{equation}
where $X(0)=0$.

Local consideration allows us to suppose that we have a
gravitational force acting on a freely falling object which is
proportional to the velocity and $H$, (see Eq. \ref{dotv}), this
force may be considered as an external force caused by the
geometry.

\subsection{Non-relativistic self-force}

 Now let us consider the limit $v \ll 1$ in the relation
($\ref{tr}$) i.e. its non-relativistic limit, so we obtain
\begin{equation}
X(t)=\frac{v_0}{2H}\Big(1-e^{-2Ht}\Big), \label{trajectory}
\end{equation}
where in the flat limit ($ H\rightarrow 0$), it leads to
$X(t)=v_0t$. And also the non-relativistic limit of external force
becomes:
\begin{equation}
f_{ext}\equiv f^G_{NR}=-2m_0Hv_0e^{-2Ht} \label{F-external},
\end{equation}
where $ f^G_{NR} $ is the non-relativistic limit of the
gravitational force and $m_0$ is the rest mass (appendix B).
Substituting (\ref{F-external}) in Eq. (\ref{E16}), we obtain:
\begin{equation}
X_q(t)=\frac{v_0}{(2H)(1+2H\alpha)}\Big(1-e^{-2Ht}\Big)=
\frac{X(t)}{(1+2H\alpha)},
\label{final-trajectory}
\end{equation}
where $X(t)$ is the trajectory of a charged particle when it
undergoes only the gravitational force, however, $X_q(t)$ is its
counterpart when we consider the effect of the self-force.
\\Suppose that, we have two completely identical objects with the
same starting points and initial velocities. It is easy to see
that because of the radiation reaction of accelerated charged
object, at any time we have $X_q(t)<X(t)$, as it was expected.

\section{Conclusion}

The covariant analysis of the electromagnetic self-force has a
quite rich history and dates back to the early work of Abraham in
$1933$, then Dirac in 1938 used an other method based upon
consideration of energy-momentum conservation in flat space-time
\cite{dirac}. The generalization of Dirac's approach to curved
space was done by DeWitt and Brehme in 1960 \cite{dewitt}. The
other interesting features of the self-force can be found in the
literature (for example see \cite{poisson2}).

In the present work, we first linearized the scalar self-force in
generic background and then studied the flat background. It was
shown if we take the perturbation as (\ref{per}), the trajectory
of object will be independent of the perturbation. \\On the other
hand, in local consideration the particle acts in a way that it is
affected by the gravitational force namely the particle moving on
its geodesic can be regarded as an object being forced under the
gravitational force. Regarding this gravitational force as an
external force, we obtained the effect of electromagnetic
self-force on particle's trajectory in non-relativistic regime. We
considered an extended charged particle, so there were not any
singularities in electromagnetical potentials. Therefore it is
shown that the trajectory of particle changes from Eq.
(\ref{trajectory}) to Eq. (\ref{final-trajectory}). Note that when
we take the limit $H\rightarrow0$, we regain the classical formula
for $X_q(t)$. This means that, we could find the
corrections in the particle's trajectory, due to its charge (self-force), in de Sitter space-time in local
consideration.\\\\\\
\textbf{Acknowledgements}\\
This work was supported by a grant from Islamic Azad University,
Central Tehran Branch. The authors would like to thank Prof. M.V.
Takook for his useful comments. MRT, would like to thank Prof. Dr.
Bayram Tekin for introducing Ref. \cite{blinnikov}.

\section{Appendix}

\subsection{\emph{consideration of the equivalence principle:}}

According to the equivalence principle, we can impose the theory
of electromagnetism into general theory of relativity. The
equivalence principle tells us that all physical statements are
applicable in locally inertial frames (even if totally
accelerated) as well as they are in Minkowski space-time.
Mathematically, this can be done by taking the covariant
derivatives instead of the partial derivatives, when the metric is
symmetric. Consequently, the electromagnetic effects, will be the
same for a charged object which is at rest.
\\The self-force is opposite to the direction of acceleration,
therefore for a moving object, the external force is acting to
neutralize the self-force effects. Also for an object which is at
rest, these two forces must be equal, so we should define a
mechanical mass ($m_{mech}$) to distinguish between a resting
object and a moving one. Using Eq. (\ref{ld}), this can be
interpreted as below, where the geodesics themselves, are
displaying the four acceleration:
\begin{equation}
m_{mech}\{\frac{d^2X^\mu}{d\tau^2}+\Gamma_{\nu\rho}^\mu\frac{dX^\nu}{d\tau}\frac{dX^\rho}{d\tau}\}=F^\mu_{ext}+F^\mu_{self}.
\label{geo-a}
\end{equation}
If the electromagnetic mass is introduced by:
\begin{equation}
m_{em}a^\mu=-F_{self}^\mu. \label{m-em}
\end{equation}
then one can rewrite (\ref{geo-a}) as:
\begin{equation}
(m_{mech}+m_{em})a^\mu=F^\mu_{ext}. \label{m-mech+m-me}
\end{equation}
And also by letting:
\begin{equation}
m_G\equiv m_{mech}+m_{em}, \label{m-G}
\end{equation}
in which $m_G$ is the gravitational mass, for a zero mechanical
mass (i.e. for a resting object), we have:
\begin{equation}
F^\mu_{ext}=-F^\mu_{self}. \label{fe=fs}
\end{equation}
and obviously
\begin{equation}
m_G=m_{em}=m_{inertial}, \label{m-G=m-me=m-inertial}
\end{equation}
where $m_{inertial}$ is the mass for a resting object. However for
a moving object, the Newton's second law reads as:
\begin{equation}
\sum_{fields}({F_{self}}) + F_{ext}=0, \label{fe+fs}
\end{equation} where the self-forces due to the other fields are
included. In this work although we worked in the locally flat
space, but the object is inserted in a gravitational field, which
is described by the general theory of relativity \cite{lyle}.

\subsection{\emph{Classical external force}}

 From Eq. (\ref{e1}) one easily obtains:

\begin{equation}
\begin{array}{l}
\frac{dt}{d\tau} = \frac{e^{-2Ht}}{v},
\end{array}
\label{E4}
\end{equation}
where $v=\frac{d\,X}{dt}$, and also it is easy to show that
\begin{equation}
\begin{array}{l}
\frac{d^2}{d\tau^2}X = -2H\frac{e^{-2Ht}}{v}e^{-2Ht}.
\end{array}
\label{E5}
\end{equation} Writing
$\frac{d^2}{d\tau^2}=\frac{d^2t}{d\tau^2}\frac{d}{dt}+(\frac{dt}{d\tau})^2\frac{d^2}{dt^2}$,
the Eq. (\ref{E5}) leads to a nonlinear differential equation with
respect to $v$ as follows:
\begin{equation}
\begin{array}{l}
-3He^{-2Ht}v+\frac{e^{-4Ht}}{v^2}\dot v+2H\frac{e^{-4Ht}}{v}=0.
\end{array}
\label{E7}
\end{equation}
This equation can be simplified to a brief expression as:
\begin{equation}
\begin{array}{l}
\dot v = -2Hv(1-\frac{3}{2}{v^2}e^{2Ht}).
\end{array}
\label{E8}
\end{equation}
In order to obtain an expression for the classical geometrical
force we start from \begin{equation}
f_{ext}=\frac{d}{d\,t}mv,\end{equation} where $m$ is relativistic
mass. So up to second order in $v$, we obtain
$\dot{m}=m_0v\dot{v}$, so it follows:
\begin{equation}
f_{ext}=m_0v^2\dot{v}+m_0\Big(1+\frac{1}{2}v^2\Big)\dot{v}.
\label{fext}\end{equation} Equation (\ref{fext}) together with
(\ref{E8}) results in: $f_{ext}=-2m_0Hv$. Following calculations
are important when we impose the effect of electromagnetic
self-force,
\begin{equation}
\begin{array}{l}
\dot v=a=\frac{f_{ext}}{m}+\alpha \dot a.
\end{array}
\label{E12}
\end{equation}
Substituting $\dot a$ in the right hand side of the Eq.
(\ref{E12}), results in:
\begin{equation}
\begin{array}{l}
a = \frac{f_{ext}}{m}+\alpha(\frac{\dot f_{ext}}{m}+\alpha\ddot
a),
\end{array}
\label{E13}
\end{equation}
and
\begin{equation}
\begin{array}{l}
ma=f_{ext}+\alpha\dot f_{ext}+\alpha^2\ddot
f_{ext}+\alpha^3f^{(3)}_{ext}+\ldots,
\end{array}
\label{E14}
\end{equation}
or we can write:
\begin{equation}
f_{self}=\sum^{\infty}_{n=1}\alpha^n(\frac{d^n}{dt^n})f_{ext}.
\end{equation}


\begin{thebibliography}{0}
\bibitem{data}  The cosmological data indicate
that for the current state of the universe we have $H_0=2.5\times
10^{-18} s^{-1}$, and $T=13.75$ billion years or $T=4.3362\times
10^{17} s$, which is the estimated age of the universe. Therefore,
for the current state we can write $H_0T\sim 1$; see also: M.R.
Tanhayi, M. Fathi, M.V. Takook, "\emph{Observable Quantities in
Weyl Gravity}" Mod. Phys. Lett. A 26, 2403 (2011) and Ref.s
therein.
\end{thebibliography}
\end{document}